\documentclass[final]{article}

\usepackage[american]{babel}
\usepackage[latin1]{inputenc}
\usepackage[T1]{fontenc}

\usepackage{amssymb}
\usepackage{amsmath}
\usepackage{bbm}

\usepackage{authblk}

\newcommand{\be}{\begin{equation}}
\newcommand{\ee}{\end{equation}}

\newcommand{\cecch}{\kappa}
\newcommand{\oehs}{\xi}
\newcommand{\idmat}{\mathbbm 1}
\newcommand{\inpr}[2]{\langle #1 | #2 \rangle}
\newcommand{\inprm}[2]{\langle #1 | \bar M_1 | #2 \rangle}
\newcommand{\ket}[1]{|#1\rangle}
\newcommand{\perm}{T}
\newcommand{\metaop}{\mathcal O}
\newcommand{\chan}{\mathcal E}
\newcommand{\hilb}{\mathcal H}
\newcommand{\oza}{\mathcal A}

\DeclareMathOperator{\tr}{tr}

\title{Error-disturbance relations for finite dimensional systems}
\date{\vspace{-5ex}} %Removes date
\author{A. C. Ipsen\thanks{asgercro@fys.ku.dk}}
\affil{The Niels Bohr Institute, University of Copenhagen\\
  Blegdamsvej 17, DK-2100 Copenhagen \O , Denmark.}

\begin{document}

\maketitle

\begin{abstract}
We propose an error-disturbance relation for general observables on 
finite dimensional Hilbert spaces based on operational notions of
error and disturbance. For two-dimensional systems we derive tight
inequalities expressing the trade-off between accuracy and disturbance.
\end{abstract}

\section{Introduction}
\label{sec:intro}

The uncertainty principle can broadly be understood in two different
ways: (a) as the impossibility of preparing a state 
such that two non-commuting observables $A$ and $\bar A$ are both sharply defined, or (b)
as the fact that  measuring  $A$ will disturb any subsequent attempt to measure  $\bar A$. 
The Robertson-Sch\"odinger inequality,
\be
  \sigma_A \sigma_{\bar A} \geq \left|\frac 1 2 \langle [A,\bar A] \rangle \right|,  
  \quad\sigma_{\metaop}^2 := \langle (\metaop-\langle \metaop \rangle)^2 \rangle,
  \label{eq:RS-ineq}
\ee
is the usual textbook example of the uncertainty principle in the first sense, whereas
there does not seem to be a broad consensus of how to express (b) in general.
A third possible formulation is that (b') it is impossible to measure both $A$ and $B$
simultaneously. There is a certain sense in which (b) and (b') 
can be regarded as equivalent\footnote{Of course, quantitative 
  expressions of (b) and (b') will not necessarily be identical.}.
Intuitively, given any apparatus capable of measuring $A$ without disturbing $\bar A$, 
it is clear that one
could construct a apparatus for the joint measurement of $A$ and $\bar A$. Conversely, if one
could measure both $A$ and $\bar A$, one could bring the system back to the appropriate eigenstate
of $\bar A$ after the measurement, thus effectively measuring $A$ without disturbing $\bar A$.
See Appendix \ref{sec:disturb} for a quantitative discussion.

We will use the term error-disturbance relation to denote an inequality expressing 
(b) or (b'). For simplicity, we will mainly focus on the (b') formulation. 
Consider two non-degenerate observables
on a $N < \infty$ dimensional Hilbert space $\hilb$. Let us write them as
\be
  A = \sum_{a=1}^N \lambda_a P_a,\qquad \bar A = \sum_{a=1}^N\bar\lambda_a \bar P_a,
\ee
where 
\be
  P_a = |a\rangle\langle a|, \qquad \bar P_a = |\bar a\rangle\langle \bar a|
\ee
are projection operators. Now a measurement of $A$, say, will yield
an eigenvalue $\lambda$, but that is clearly equivalent to specifying the integer $a$
such that $\lambda_a = \lambda$ (here we use that $A$ is non-degenerate). We will thus
forget about the eigenvalues in the following, and focus on the projection 
operators\footnote{This is similar to the approach
  in e.g. the case of entropic uncertainty relations\cite{De83,MU88}.}
(equivalently the eigenbases).

We describe the apparatus for (approximate) joint measurement of $A$ and $\bar A$ by 
a Positive Operator Valued Measure (POVM). In more detail, we consider
$N^2$ positive (positive will always mean positive semidefinite for operators) operators $F_{ab}$
such that 
\be
  \sum_{a,b}^N F_{ab} = \idmat,
\ee
and define the probability of getting output $(a,b)$ given a state $\rho$ to be
\be
  \mathcal P_{\rho}(a,b) := \tr (F_{ab}\rho).
\ee
An output $(a,b)$ should intuitively be understood as indicating the joint measurement result
$A = \lambda_a$ and $\bar A = \bar \lambda_b$. 

We now want to quantify how good the apparatus is at measuring $A$ and $\bar A$.
Let us define two POVMs corresponding to the first and second output of the apparatus:
\be
  M_a := \sum_{b}F_{ab}, \qquad \bar M_{a} := \sum_b F_{ba}.
\ee
We consider the apparatus to be `good' if  these should are, respectively, close to the projective
POVMs $P$ and $\bar P$. We thus need to introduce a metric, $d(M,P)$,
to define what we mean by `close'. We will consider two different possibilities in the following.

The most basic way of testing the apparatus one could think of is to prepare the system
in an eigenstate, $\rho = P_a$, say, and then checking how often the apparatus gives
the right answer. This procedure naturally leads to the definition
\be
  d_c(M,P) := \sup_{a=1,\ldots,N}(1-\tr(M_aP_a)).
  % =\sup_a(1-\langle a|M_a|a\rangle)
\ee
In words, $d_c(M,P)$ is the worst case error probability if the system is prepared in an
eigenstate of $A$. It is clear that
\be
  d_c(M,P) \geq 0,
\ee
and it is easy to show that $d_c(M,P) = 0$ iff $M_a = P_a$ for all $a$. We will call $d_c(M,P)$
the \emph{calibration error}, since it is similar in spirit to the one used in \cite{BLW13} for
continuous variables.

For a fixed state $\rho$, $M$ and $P$ each define a probability distribution, so 
another approach would be to compare these distributions. A natural distance measure
on probability distributions is the \emph{total variation distance}\footnote{The 
  total variation distance was also considered (in the context of approximate measurements)
  in \cite{BHL04}.}
\be
  \frac 1 2\sum_a |\tr(M_a\rho)-\tr(P_a\rho)|.
\ee
We then define the \emph{variation error} by taking the supremum over all states:
\be
  d_v(M,P) := \sup\left\{\frac 1 2\sum_a |\tr(M_a\rho)-\tr(P_a\rho)|
    \middle|\rho\ge 0,\tr\rho=1\right\}.
  \label{eq:d-v-def}
\ee
Again we have that
\be
  d_v(M,P) \geq 0,
\ee
and that $d_v(M,P) = 0$ iff $M_a = P_a$ for all $a$. We also find the relation
\be
  d_v(M,P) \ge d_c(M,P)
  \label{eq:v-geq-c}
\ee
by writing
\be
  d_c(M,P) = \sup\left\{\frac 1 2\sum_a |\tr(M_a\rho)-\tr(P_a\rho)|
    \middle|\rho=|a\rangle\langle a|,a=1,\ldots,N\right\}.
  \label{eq:d-c-sup}
\ee
Let $\mathcal F$ be the powerset of $\{1,\ldots,N\}$, then it is easy to
see that
\be
  \frac 1 2\sum_a |\tr(M_a\rho)-\tr(P_a\rho)|
    = \sup_{X\in \mathcal F}|\tr(M(X)\rho)-\tr(P(X)\rho)|,
\ee
where
\be
  M(X) = \sum_{a\in X} M_a,\qquad
  P(X) = \sum_{a\in X} P_a.
\ee
It follows that we can alternatively write $d_v$ as
\be
  d_v(M,P) = \sup_{X\in \mathcal F}\|M(X)-P(X)\|,
\ee
where $\|\cdot\|$ is the operator norm.

The choice between $d_c$ and $d_v$ will be dictated by the application of the
error-disturbance relation. The calibration error has a very simple operational interpretation,
while the variation error treats the input states in a more uniform way (i.e. without singling
out the eigenstates). We will see in Section \ref{sec:qubit} that already for the qubit,
the two error metrics lead to inequivalent optimal joint measurement schemes. 

We are now ready to write down the general form of our error-disturbance relation.
Denote the errors by
\begin{align}
  \epsilon_{\beta} &= d_\beta(M,P), &\bar\epsilon_\beta &= d_\beta(\bar M,\bar P),\qquad
  \beta\in\{c,v\},
\end{align}
we then want to consider inequalities of the form
\be
  G_\beta(\epsilon_\beta,\bar \epsilon_\beta) \geq B_\beta(P,\bar P),
  \label{eq:ed-relation}
\ee
valid for all POVMs $F$. We remark that this is a \emph{state independent} relation.
The expression $G(\epsilon,\bar\epsilon)$ is meant to quantify the total error of the
apparatus, and there seems to be many valid choices for functional form of $G$.
The Roberson-Sch\"odinger-like choice $G(\epsilon,\bar\epsilon) = \epsilon\bar\epsilon$,
however,
is \emph{not} a good one, since there is an apparatus such that $\epsilon =0$
and $\bar\epsilon$ is finite (the errors are bounded in our case) ruling out any
non-trivial bound in \eqref{eq:ed-relation}.

We will focus on the case $G(\epsilon,\bar\epsilon) := \epsilon+\bar\epsilon$, see
Appendix \ref{sec:G-choice} for a heuristic motivation of this choice.
The task thus becomes to determine the strongest possible bound\footnote{Formally we
  want to set $B(P,\bar P) = \inf_F\{\epsilon+\bar\epsilon\}$, but that is clearly only
  useful if we know how to compute the infimum.}
$B(P,\bar P)$ making 
\be
  \epsilon_\beta + \bar\epsilon_\beta \geq B_\beta(P,\bar P)
  \label{eq:ed-relation2}
\ee
valid. Note that we will sometimes omit the index $\beta$ for brevity.

In Appendix \ref{sec:disturb} we show how \eqref{eq:ed-relation2} can also
be interpreted as
\be
  \epsilon + \bar\eta \geq B(P,\bar P),
\ee
where $\epsilon$ is the error of a quantum instrument measuring $A$ and $\bar\eta$
is the disturbance of $\bar A$ incurred by the instrument.

\subsection{Results}
In Section \ref{sec:qubit} we will compute the optimal value of $B_\beta(P,\bar P)$ 
for the qubit. Specifically, we find the \emph{tight} relations 
\be
  \epsilon_v+\bar\epsilon_v \geq \sin\left(\frac \pi 4 +\theta\right)-\frac{\sqrt 2}{2}
  \label{eq:ev-tight}
\ee
and
\be
  \epsilon_c+\bar\epsilon_c \geq 2\sin^2 \frac \theta 2.
\ee
Here $0 \leq 2\theta \leq \frac \pi 2$ is the angle between $|1\rangle$ and $|\bar 1\rangle$ 
on the Bloch sphere.

In Section \ref{sec:MUB} we consider the case where the bases $|a\rangle$ and 
$|\bar a\rangle$ are mutually unbiased. Using some results from Appendix \ref{sec:gen-bound},
we derive
\be
  \epsilon_c+\bar\epsilon_c \geq 2\left(\frac 1 7\left(4\sqrt 2 -5\right)\right)^2
\ee
for $N=3$, and
\be
  \epsilon_c+\bar\epsilon_c \geq 2\left(\frac{1}{31}\left(4\sqrt{7} -9\right)\right)^2
\ee
for $N=5$. By \eqref{eq:v-geq-c} these also hold for the variation error.

After the completion of this manuscript earlier work by Bush and Heinosaari \cite{BH08} came
to our attention. Their results overlap with our results for the qubit. In particular
they also derive \eqref{eq:ev-tight}.

\subsection{Related work}
The formulation of our error-disturbance relation is closest in spirit to the one
by Bush, Lahti, Pearson and Werner \cite{We04,BP07,BLW13} for canonical position and
momentum operators, but a direct comparison is not possible, since we work with 
finite dimensional systems. 
Here we will briefly compare our proposed inequality with two other proposals.

In order to discuss Ozawa's uncertainty relation \cite{Oz03,Oz04,Ha04,Br13}, we need
to introduce an auxiliary Hilbert space $\hilb_{aux}$ and a fixed state $\rho_{aux}$ on
$\hilb_{aux}$. The measurement apparatus is then described by two \emph{commuting}
observables\footnote{One usually writes 
  $\oza = U^\dagger (\idmat_\hilb \otimes \oza_{aux}) U$ (and similarly for $\bar \oza$), 
  where $\oza_{aux}$ is an observable
  on $\hilb_{aux}$ and $U$ is a unitary describing some interaction between the system of interest
  and the auxiliary system.}
$\oza,$ $\bar\oza$ on $\hilb\otimes\hilb_{aux}$. Define the errors (the $O$ is part of the name, 
not an index)
\be
  \epsilon^2_{O,\rho} = \langle (\oza-A\otimes\idmat_{aux})^2 \rangle_{\rho\otimes\rho_{aux}},\qquad
  \bar\epsilon^2_{O,\rho} = \langle (\bar\oza-\bar A\otimes\idmat_{aux})^2
    \rangle_{\rho\otimes\rho_{aux}},
\ee
and the standard deviations
\be
  \sigma^2_\rho = \langle (A-\langle A \rangle_\rho)^2 \rangle_\rho,\qquad
  \bar\sigma^2_\rho = \langle (\bar A-\langle \bar A \rangle_\rho)^2 \rangle_\rho.
\ee
One can then derive the following error-disturbance relation\footnote{In Ref. \cite{Ha04} Hall derives 
  a very similar inequality, but with $\sigma$ ($\bar\sigma$) defined in terms of $\oza$ ($\bar\oza$).
  A tight variant of \eqref{eq:ozawa} is given by Branciard \cite{Br13}. See also
  Weston et al. \cite{WHPWP13} and Lu et al. \cite{LYFO13} for related inequalities.}\cite{Oz04}
\be
  \epsilon_{O,\rho}\bar\epsilon_{O,\rho}+\epsilon_{O,\rho}\bar\sigma_\rho
    +\sigma_\rho\bar\epsilon_{O,\rho}
    \geq \frac 1 2 |\langle [A,\bar A] \rangle_\rho |.
  \label{eq:ozawa}
\ee

Another relation is due to Hofmann\cite{Ho03}. Consider a POVM (the $H$ is not an index)
\be
  \sum_m F_{H,m} = \idmat,
\ee
where $m$ ranges over some set of measurement outcomes (the number of outcomes does \emph{not}
have to be related to the dimension $N$ of $\hilb$), and introduce the errors
\be
  \epsilon_{H,m}^2 = \langle (A-\langle A \rangle_{\rho_m})^2 \rangle_{\rho_m},\qquad
  \bar\epsilon_{H,m}^2 = \langle (\bar A-\langle \bar A \rangle_{\rho_m})^2 \rangle_{\rho_m}.
\ee
Here $\rho_m$ is the ``retrodictive'' state corresponding to the measurement outcome $m$,
explicitly
\be
  \rho_m = \frac{F_{H,m}}{\tr [F_{H,m}]}.
\ee
We then have the  relation\cite{Ho03} (see also \cite{DN13})
\be
  \epsilon_{H,m}\bar\epsilon_{H,m} \geq \frac 1 2 |\langle [A,\bar A] \rangle_{\rho_m}|.
  \label{eq:Hofmann}
\ee

An obvious difference between these error-disturbance relations and ours is that here
the spectra of $A$ and $\bar A$ plays a role, while in \eqref{eq:ed-relation2} only
the spectral bases appear. A more substantial difference is that
the RHS of \eqref{eq:ozawa} and \eqref{eq:Hofmann} depends on a state (the (pre-measurement) 
system state in the Ozama case and the retrodictive state in the Hofmann case). For the Hofmann
relation the RHS further depends on the details of the measurement apparatus. In contrast,
the RHS of \eqref{eq:ed-relation2} only depends on the operators we want to measure. This
means that our relation is non-trivial as long as $A$ and $\bar A$ do not commute\footnote{It
  follows from the results in Appendix \ref{sec:gen-bound} that $B(P,\bar P)$ can be taken to
  non-zero when $A$ and $\bar A$ do not commute.},
independent of the system state and apparatus, which is not in general the case for
\eqref{eq:ozawa} and \eqref{eq:Hofmann}.

\section{The qubit}
\label{sec:qubit}
We will now consider the simplest non-trivial case, that is $N=2$.
We will see that it is possible to determine the optimal bound $B$ in \eqref{eq:ed-relation2}.
Even for qubit measurements, it takes 12 real parameters to specify the $F$s,
so computing the infimum directly would be demanding. Fortunately, almost all
the degrees of freedom can be eliminated using symmetry. To see how this works,
let us choose the axis on the Bloch sphere such that
\be
  P_1 = \frac 1 2\left(\idmat+(\cos\theta)\sigma_z+(\sin\theta)\sigma_x\right),
\ee
and
\be
  \bar P_1 = \frac 1 2\left(\idmat+(\cos\theta)\sigma_z-(\sin\theta)\sigma_x\right),
\ee
where $0\leq\theta\leq\frac \pi 4$ (we might have to relabel, say, 
$|1\rangle\leftrightarrow |2\rangle$
to allow this). We see that a $180^\circ$ rotation along the $z$-axis will exchange the
barred and unbarred basis. If we combine the rotation with a relabeling 
$|1\rangle \leftrightarrow |\bar 1\rangle$ and $|2\rangle \leftrightarrow |\bar 2\rangle$
we thus get a new measurement scheme.
Taking the average of the original and the rotated measurement scheme, as in Appendix
\ref{sec:G-choice}, we get a measurement scheme which is symmetric, and at least as good.
We will now make this more precise.

Let $S$ be an unitary or antiunitary operator on $\hilb$. % involution (i.e. $S^2=\idmat$).
Then 
\be
  \rho \mapsto S\rho S^\dagger
\ee
maps the space of states bijectively onto itself. It follows that
(here $(S^\dagger MS)_a = S^\dagger M_a S$, etc.)
\be
  d_\beta(S^\dagger MS,P) = d_\beta(M,SPS^\dagger).
  \label{eq:d-invariance}
\ee
If we now take $S=\sigma_z$, and set (it is clear that $F'$ is a POVM)
\begin{align}
  F'_{11} &= \sigma_z F_{11} \sigma_z,
    & F'_{12} &= \sigma_z F_{21} \sigma_z,\\
  F'_{21} &= \sigma_z F_{12} \sigma_z,
    & F'_{22} &= \sigma_z F_{22} \sigma_z,
\end{align}
then
\be
  \epsilon' = d(M',P) = d(\bar M,\bar P),\qquad
  \bar\epsilon' = d(\bar M',\bar P) = d(M,P).
\ee
By taking the average of $F$ and $F'$,
we see that the infimum of $G(\epsilon,\bar\epsilon) = \epsilon+\bar\epsilon$ is achieved 
by $F$s satisfying
\begin{subequations}\label{eq:F-sym}
\begin{align}
  F_{11} &= \sigma_z F_{11} \sigma_z = \sigma_y F_{22}\sigma_y,
    & F_{12} &= \sigma_z F_{21} \sigma_z = \sigma_y F_{21}\sigma_y,\\
  F_{21} &= \sigma_z F_{12} \sigma_z = \sigma_y F_{12}\sigma_y,
    & F_{22} &= \sigma_z F_{22} \sigma_z = \sigma_y F_{11}\sigma_y.
\end{align}
The second equal sign in the equations follows from similar considerations with
$S = \sigma_y$. Finally, by letting $S$ be the antilinear 
operator\footnote{With the standard choice for the Pauli matrices this is
  simply complex conjugation.}
corresponding to reflection in the $x$-$z$ plane, we find that we can further
restrict to $F$s with no $\sigma_y$ components, i.e. 
\be
  \tr(F_{ab}\sigma_y) = 0\quad\text{[for all $a$ and $b$]}.
  \label{eq:no-sigma-y}
\ee
\end{subequations}

In this symmetric subspace $\epsilon = \bar\epsilon$, so the task becomes to find
the smallest possible
value of $\epsilon$. We will now do this for each of the error metrics.

\subsection{Variation error}
By \eqref{eq:no-sigma-y}, we can parametrize $M_1$ as
\be
  M_1 = \frac 1 2(\idmat+(\cos\theta+2\epsilon\cos \chi)\sigma_z
  +(\sin\theta+2\epsilon\sin\chi)\sigma_x).
\ee
The use of $\epsilon$ is consistent, since 
a simple calculation shows that
\be
  d_v(M,P) = \epsilon.
\ee
%\be
%  \rho = \frac 1 2\left(1+(\cos\phi)\sigma_z+(\sin\phi)\sigma_x\right)
%\ee
%\be
%  \tr(M_1\rho)-\tr(P_1\rho) = \epsilon(\cos\phi)(\cos\chi)+\epsilon(\sin\phi)(\sin\chi)
%\ee
Note that the symmetries \eqref{eq:F-sym} fix the other three marginals,
but the $F$s are not completely fixed. We parametrize the freedom by $a$, setting
\be
  F_{11} = aP_+ + (a-\cecch)P_-,
\ee
and consequently
\be
  F_{12} = \frac 1 2 ([1+\cecch-2a]\idmat
  +[\sin\theta+2\epsilon\sin\chi]\sigma_x),
\ee
%\be
%  F_{2\bar 2} = (a-\cecch)P_+ + (b+\cecch)P_-
%\ee
in terms of the projections
\be
  P_{\pm} = \frac 1 2 (\idmat \pm \sigma_z),
\ee
and with 
\be
  \cecch = \cos\theta+2\epsilon\cos\chi.
  \label{eq:ccech}
\ee
It is clear that $F_{11}$  is positive iff
$a\geq 0$ and $a-\cecch \geq 0$. 

We assume that we have found a point $(\epsilon,\chi,a)$ such that $\epsilon$ is 
minimal and the $F$s are positive. Note that $F_{11}$ ($F_{12}$) is
positive iff $F_{22}$ ($F_{21}$) is positive.
We will assume that
$\cecch > 0$  for now. Then we can set $a=\cecch$, since making $a$ smaller makes
$F_{12}$ more positive, thus
\be
  F_{11} = \cecch P_+,
\ee
and
\be
  F_{12} = \frac 1 2([1-\cecch]\idmat+[\sin\theta+2\epsilon\sin\chi]\sigma_x).
\ee
If $F_{12}(\epsilon,\chi)$ was strictly positive, then, by continuity, 
one could find a $\epsilon'<\epsilon$
such that $F_{12}(\epsilon',\chi)$ would still be positive, thus this cannot
be the case. We conclude
that at least one of the eigenvalues of $F_{12}$ must be zero, or
\be
  \det F_{12} = 0.
  \label{eq:det0}
\ee
If now 
\be
  \frac{\partial\det F_{12}}{\partial \chi} \neq 0
\ee
then we could find a $\chi'$ near $\chi$ such that both the eigenvalues would be
strictly positive (with the same $\epsilon$), a contradiction by the above argument, thus
\be
  \frac{\partial\det F_{12}}{\partial \chi} = 0.
  \label{eq:ddet0}
\ee

It is now straight forward to find the solutions to \eqref{eq:det0} and \eqref{eq:ddet0}.
The one with the smallest $\epsilon$, amongst those satisfying $\tr F_{12} \geq 0$, 
has\footnote{Incidentally, the POVM \eqref{eq:c-povm} also solves the equations, but with
  $\epsilon = \sin(\theta/2)$, which is worse than \eqref{eq:e-optimal-v}.}
\be
  \chi_{\text{optimal},v} = \frac{5\pi}{4}
\ee
and
\be
  \epsilon_{\text{optimal},v}  = \frac 1 2 \left(\sin\left(\frac \pi 4 +\theta\right)-\frac{\sqrt 2}{2}\right).
  \label{eq:e-optimal-v}
\ee
For $\cecch \leq 0$ it follows directly from \eqref{eq:ccech} that
\be
  \epsilon \geq \frac{\cos\theta}{2},
\ee
which is worse than \eqref{eq:e-optimal-v}.
We thus have an optimal bound
%\be
%  G(d_v(M,P),d_v(\bar M,\bar P)) 
%    \geq G\left(\epsilon_{\text{optimal},v},\epsilon_{\text{optimal},v}\right),
%\ee
\be
  d_v(M,P) + d_v(\bar M,\bar P) 
    \geq 2\epsilon_{\text{optimal},v} = \sin\left(\frac \pi 4 +\theta\right)-\frac{\sqrt 2}{2}
\ee
Explicitly, the POVM takes the form
\begin{subequations}
\begin{align}
  F_{11} &= \frac 1 2(1+\cos\theta-\sin\theta)P_+
  & F_{12} &= \frac 1 4(1-\cos\theta+\sin\theta)(\idmat+\sigma_x),\\
  F_{21} &= \frac 1 4(1-\cos\theta+\sin\theta)(\idmat-\sigma_x)
  & F_{22} &= \frac 1 2(1+\cos\theta-\sin\theta)P_-.
\end{align}
\end{subequations}

\subsection{Calibration error}
The calculations in the case of the calibration error are similar to those of
the previous section, so we give fewer details.
We set 
\be
  M_1 = \frac 1 2(\idmat + \oehs \sigma_z+h\sigma_x),
\ee
with 
\be
  \oehs = \frac{1-2\epsilon-h\sin\theta}{\cos\theta},
\ee
and, for $\oehs > 0$,
\be
  F_{11} = \oehs P_+.
\ee
%\be
%  d_c(M,P) = \epsilon
%\ee
The analogs of \eqref{eq:det0} and \eqref{eq:ddet0} hold (with $h$ instead of $\chi$),
and we find\footnote{At the point $\theta = \pi/4$ the symmetry of the 
  problem is enhanced, and the solution is optimal for any $h \in [0,1]$.}
\be
  h_{\text{optimal},c} = 0,
\ee
and
\be
  \epsilon_{\text{optimal},c} = \sin^2 \frac \theta 2.
\ee
The case of $\oehs \leq 0$ is also easily treated, and is seen to lead to
suboptimal solutions (this is also intuitive, since we would have $F_{11}\propto P_-$).
As before, we deduce the tight bound
\be
  d_c(M,P)+d_c(\bar M,\bar P) 
    \geq 2\epsilon_{\text{optimal},c} = 2\sin^2 \frac \theta 2.
\ee
The optimal POVM is just a projective measurement along the $z$-axis, i.e.
\begin{subequations}\label{eq:c-povm}
\begin{align}
  F_{11} &= P_+ & F_{12} &= 0,\\
  F_{21} &= 0   & F_{22} &= P_-.
\end{align}
\end{subequations}

%\begin{subequations}
%\begin{align}
%  F_{1\bar 1} &= (1-h)P_+ & F_{1\bar 2} &= \frac{h}{2}(\idmat+\sigma_x)\\
%  F_{2\bar 1} &= \frac{h}{2}(\idmat-\sigma_x) & F_{2\bar 2} &= (1-h)P_-
%\end{align}
%\end{subequations}

\section{Mutually unbiased bases}
\label{sec:MUB}
Two bases are said to be mutually unbiased if
\be
  |\inpr{a}{\bar b}|^2 = \frac 1 N
  \label{eq:MUB}
\ee
for all $a$ and $b$. Here we will show that the symmetry condition \eqref{eq:E-symm},
which basically says that given a POVM one can construct a new one with $\epsilon$
and $\bar\epsilon$ interchanged (see Appendix \ref{sec:G-choice}), 
holds for unbiased bases in three and five dimensions. This allows us to deduce 
explicit error-disturbance relations using the results of Appendix \ref{sec:gen-bound}.

Given a basis
$|a\rangle$ we can construct a new basis $|\bar a\rangle$ by Fourier transformations,
\be
  \langle a \ket{\bar b} = \frac{1}{\sqrt{N}}e^{i\frac{2\pi ab}{N}},
  \label{eq:fourier}
\ee
and it is clear that \eqref{eq:MUB} is satisfied. We will first show that \eqref{eq:E-symm}
holds when the bases are related as in \eqref{eq:fourier}. Define the unitary operator $U$ by
\be
  \ket{\bar a} = U\ket{a},
\ee
%\be
%  \langle a|U\ket{b} = e^{i\frac{2\pi ab}{N}}
%\ee
then it follows from \eqref{eq:fourier} that
\be
  U^2\ket{a} = U\ket{\bar a} = \ket{-a},
\ee
where $-a$ is understood modulo $N$. Given a POVM $F_{ab}$, define
a new one by
\be
  F'_{ab} = U^\dagger F_{-b,a} U.
\ee
Using the invariance of $d$, Eq. \eqref{eq:d-invariance}, we have
\be
  \epsilon' = d(M',P) = d(\bar M,\bar P) = \bar \epsilon,
\ee
and
\be
  \bar\epsilon' = d(\bar M',\bar P) = d(\{M_{-a}\}_a,\{P_{-a}\}_a) = \epsilon,
\ee
which demonstrates \eqref{eq:E-symm}.

Given a pair of bases that satisfy \eqref{eq:MUB} they will not generally be
related by \eqref{eq:fourier}, but sometimes they can be brought into that form
by symmetry operations. Specifically, it is clear that
permuting the basis vectors and changing their phases will leave $\epsilon$ and $\bar\epsilon$
invariant when combined with the appropriate permutation of the $F$s.
More formally, let the $N\times N$
matrix $H$ be given by (the reason for the normalization will become clear)
\be
  H_{ab} = \sqrt{N}\inpr{a}{\bar b}.
\ee
We now want to know whether we can find diagonal unitary matrices $D_1,D_2$ and
permutation matrices $\perm_1,\perm_2$ such that
\be
  (D_1\perm_1 H \perm_2 D_2)_{ab} = e^{i\frac{2\pi ab}{N}}.
  \label{eq:had-fourier}
\ee
This problem has been studied. In fact, it is clear that
\be
  |H_{ab}| = 1,\qquad HH^\dagger = N\idmat,
\ee
which makes $H$ a complex
Hadamard matrix (see e.g. \cite{TZ06}), and the equivalence modulo permutations and
phases\footnote{That is $H\sim H'$ iff $H' = D_1\perm_1 H \perm_2 D_2$ for some $D$s and $\perm$s.}
appearing in \eqref{eq:had-fourier} is exactly the equivalence that Hadamard
matrices are classified up to. For $N=3$ and $N=5$ it is known \cite{Ha96,TZ06}
that one can always find $D$s and $\perm$s such that \eqref{eq:had-fourier} is
satisfied. In other words, unbiased bases in three and five dimensions are essentially
Fourier pairs (in the sense of \eqref{eq:fourier}).

When \eqref{eq:E-symm} holds, we can restrict attention to POVMs with
 $\epsilon = \bar\epsilon$ as in Section \ref{sec:qubit}.
Using the results of Appendix \ref{sec:gen-bound}, we then derive the following bound
for mutually unbiased bases with $N=3$,
\be
  \epsilon_c+\bar\epsilon_c \geq 2\left(\frac 1 7\left(4\sqrt 2 -5\right)\right)^2,
  \label{eq:MUB-ed-3}
\ee
%\be
%  \epsilon+\bar\epsilon \geq 2\left(\frac{1}{17}\left(\sqrt{66} -7\right)\right)^2
%\ee
and for $N=5$ we have
\be
  \epsilon_c+\bar\epsilon_c \geq 2\left(\frac{1}{31}\left(4\sqrt{7} -9\right)\right)^2.
  \label{eq:MUB-ed-5}
\ee

\section{Further work}
The estimates in Appendix \ref{sec:gen-bound} are not optimal and possibly they are quite far
from optimal, so a reasonable next step would be to attempt improve them.
This would immediately lead to sharper bound in \eqref{eq:MUB-ed-3} and \eqref{eq:MUB-ed-5}.

I would also be interesting to understand under which conditions (if any) the symmetry
condition \eqref{eq:E-symm} fails.

While this manuscript was being finished, a preprint appeared introducing a information-theoretic
error-disturbance relation\cite{BHOW13}. Their approach seems similar to ours in some ways, and
it would be interesting to explore the exact relationship.

\section{Acknowledgment}
We thank C.~Branciard for useful comments.
Support from the ERC-Advanced grant 291092
``Exploring the Quantum Universe'' (EQU) is acknowledged.

\appendix

\section{The choice of $G$}
\label{sec:G-choice}
We will demand that $G$ should be monotone, 
\be
  G(\epsilon',\bar\epsilon') \leq G(\epsilon,\bar\epsilon)
  \label{eq:G-mon}
\ee
when $\epsilon'\leq\epsilon$ and $\bar\epsilon'\leq\bar\epsilon$,
and symmetric, 
\be
  G(\epsilon,\bar\epsilon) = G(\bar\epsilon,\epsilon).
  \label{eq:G-symm}
\ee
We show that, given a symmetry assumption on the space of measurements 
(Eq. \eqref{eq:E-symm}), the choice
\be
  G(\epsilon,\bar\epsilon) = \epsilon+\bar\epsilon
\ee
is the strongest among the convex $G$.

First let us note that given two POVMs $F^{(1)}$ and $F^{(2)}$, we can
form a new one by convex combination:
\be
  F^{(3)}_{ab} = \alpha F^{(1)}_{ab}+(1-\alpha)F^{(2)}_{ab},\qquad
  0 \leq \alpha \leq 1.
\ee
% From \eqref{eq:d-c-sup} and \eqref{eq:d-v-def} it is easy to see that the
% metrics are convex in the sense that
% \be
%   d(M^{(3)},P) \leq \alpha d(M^{(1)},P)+(1-\alpha)d(M^{(2)},P)
%   \label{eq:d-convex}
% \ee
% for both $d=d_c$ and $d=d_v$.
From \eqref{eq:d-c-sup} and \eqref{eq:d-v-def} it is easy to see that the
process of combining the POVMs does not introduce additional error, in the
sense that
\be
   \epsilon^{(3)} \leq \alpha \epsilon^{(1)}+(1-\alpha)\epsilon^{(2)},
   \label{eq:d-convex}
\ee
with 
\be
  \epsilon^{(i)} = d(M^{(i)},P).
\ee
An identical statement holds for $\bar\epsilon$, of course. It would be natural
to choose $G$ such that this property also holds for the combined error 
$G(\epsilon,\bar\epsilon)$. This is insured if $G$ is convex,
\be
  G(\alpha\epsilon +(1-\alpha)\epsilon',\alpha\bar\epsilon+(1-\alpha)\bar\epsilon')
    \leq \alpha G(\epsilon,\bar\epsilon) + (1-\alpha)G(\epsilon',\bar\epsilon'),
  \label{eq:G-convex}
\ee
as we then indeed have
\be
  G(\epsilon^{(3)},\bar\epsilon^{(3)}) \leq \alpha G(\epsilon^{(1)},\bar\epsilon^{(1)})
  +(1-\alpha)G(\epsilon^{(2)},\bar\epsilon^{(2)}).
\ee

For a given pair of bases, let $E(\epsilon,\bar\epsilon)$ abbreviate
``there exists a POVM with $\epsilon=d(M,P)$ and $\bar\epsilon=d(\bar M,\bar P)$'',
and set
\be
  2\epsilon_{\text{inf}} = \inf\{\epsilon+\bar\epsilon|E(\epsilon,\bar\epsilon)\}.
\ee
Let us assume the following symmetry
\be
  E(\epsilon,\bar\epsilon) \iff E(\bar\epsilon,\epsilon),
  \label{eq:E-symm}
\ee
which is shown to hold for the qubit in Section \ref{sec:qubit} and
for mutually unbiased bases (in 3 and 5 dimension) in Section \ref{sec:MUB}. 
Using \eqref{eq:d-convex} we then have\footnote{Here we also use that
  one can always modify a given POVM to make $\epsilon$ and $\bar\epsilon$ 
  bigger (within the relevant limits).}
\be
  E(\epsilon,\bar\epsilon) \Rightarrow 
    E\left(\frac 1 2(\epsilon+\bar\epsilon),\frac 1 2(\epsilon+\bar\epsilon)\right).
\ee
We thus have
\be
  \epsilon_{\text{inf}} = \inf\{\epsilon|E(\epsilon,\epsilon)\}.
\ee
Consider now a convex function $G$ as just discussed, and set
\be
  G_{\text{inf}} = \inf\{G(\epsilon,\bar\epsilon)|E(\epsilon,\bar\epsilon)\}.
\ee
If 
\be
  \epsilon+\bar\epsilon \geq 2\epsilon_{\text{inf}}
  \label{eq:epsineq}
\ee 
we have
(the first inequality is by \eqref{eq:G-mon}, \eqref{eq:G-symm} and \eqref{eq:G-convex})
\begin{align}
  G(\epsilon,\bar\epsilon) 
    &\geq G\left(\frac 1 2(\epsilon+\bar\epsilon),\frac 1 2(\epsilon+\bar\epsilon)\right)\\
    &\geq G(\epsilon_{\text{inf}},\epsilon_{\text{inf}})\\
    &\geq G_{\text{inf}}.
\end{align}
In other words, the inequality \eqref{eq:epsineq} implies any inequality
(of the form \eqref{eq:ed-relation}) with a convex $G$, assuming \eqref{eq:E-symm}.

\section{The disturbance caused by measurement}
\label{sec:disturb}
To discuss the disturbance incurred by an approximate measurement of 
$A$, we need to keep track of the post-measurement state. The correct
notion is that of a \emph{quantum instrument} (see e.g. \cite{Da76}).
A quantum instrument can though of as a quantum channel that 
also has a classical output (i.e. the outcome of the measurement).
In more detail, we associate a completely positive\footnote{A map is positive
  iff it sends positive operators to positive operators. It is completely positive
  iff the induced (linear) map on operators on $\hilb\otimes\hilb_{aux}$ (defined by
  $\rho\otimes\rho_{aux} \mapsto \chan_a(\rho)\otimes\rho_{aux}$) is positive
  for all (finite dimensional) $\hilb_{aux}$.}  map
\be
  \rho \mapsto \chan_a(\rho)
\ee
to each measurement outcome $a$
which sends positive operators on $\hilb$ to positive operators on $\hilb$.
The probability of a  outcome given a state $\rho$ is
\be
  \label{eq:instr-prob}
  \mathcal P_\rho (a) = \tr[\chan_a(\rho)],
\ee
and the (normalized) post-measurement state is 
\be
  \frac{\chan_a(\rho)}{\tr[\chan_a(\rho)]}.
\ee
From \eqref{eq:instr-prob} it follows that we must have
\be
  \sum_a \tr[\chan_a(\rho)] = \tr(\rho) = 1,
\ee
for all states $\rho$ (note that the individual $\chan_a$ do not preserve trace).

If we perform a projective measurement of $\bar A$ after the measurement of $A$,
 the joint probability distribution of the two measurements is
\be
  \mathcal P_\rho (a,b)
  = \langle \bar b|\chan_a(\rho)|\bar b \rangle.
\ee
It is a well known result due to Choi and Kraus that we can find operators
$V_{ak}$ such that
\be
  \chan_a(\rho) = \sum_k V_{ak}\rho V_{ak}^\dagger,\qquad
  \sum_{ak} V_{ak}^\dagger V_{ak} = \idmat,
\ee
where $k$ ranges over some finite set.
We can thus write
\be
  \mathcal P_\rho (a,b) = \tr(F_{ab}\rho),
\ee
where 
\be
  F_{ab} = \sum_k V_{ak}^\dagger |\bar b\rangle \langle \bar b| V_{ak}.
\ee
It is easy to check that $F_{ab}$ is a POVM.

We conclude that the joint measurement
relation \eqref{eq:ed-relation2} translates to the error-disturbance 
relation (with the same bound $B(P,\bar P)$)
\be
  \epsilon + \bar\eta \geq B(P,\bar P)
  \label{eq:edr}
\ee
for the instrument
\be
  \chan_a(\rho) = \sum_k V_{ak}\rho V_{ak}^\dagger,
\ee
where
\be
  \epsilon = d(M,P),\qquad \bar\eta = d(\bar M,\bar P),
\ee
and
\be
  M_a = \sum_k V_{ak}^\dagger V_{ak},\qquad
  \bar M_b = \sum_{a,k} V_{ak}^\dagger |\bar b\rangle \langle \bar b| V_{ak}.
\ee
The interpretation of $\epsilon$ is as in Section \ref{sec:intro}, i.e.
it quantifies how much the measurement defined by the instrument $\chan_a$
deviates from the ideal (projective) measurement of $A$.
On the other hand, $\bar \eta$ is the error of a projective measurement of $\bar A$
performed after the instrument, relative to a projective measurement of $\bar A$ before
the instrument.
It is thus natural to interpret $\bar\eta$ as the disturbance of $\bar A$ caused 
by the measurement of $A$. This justifies calling \eqref{eq:edr} an error-disturbance relation.

Let us note that the notion of a quantum instrument is 
sufficiently general to include any  error correction one could perform on the
system after learning result of the $A$ measurement, so \eqref{eq:edr} holds even with
error correction.

\section{A general bound}
\label{sec:gen-bound}
Here we will derive a bound for general dimension $N\geq 2$.
We will first need some simple properties of positive operators.

Let $K \geq 0$ be an operator. If
\be
  \langle 1 |K|1\rangle,\langle 2 |K|2\rangle \leq \epsilon
\ee
then
\be
  |\langle 1|K|2\rangle| \leq  \epsilon,
\ee
while if
\be
  \langle 1 |K|1\rangle\leq 1\text{, and }\langle 2 |K|2\rangle \leq \epsilon
\ee
then
\be
  |\langle 1|K|2\rangle| \leq \sqrt {\epsilon}.
\ee
These relations can be derived by noting that 
\be 
  (\langle 1| +\alpha^*\langle 2|)K(|1\rangle +\alpha|2\rangle) \geq 0
\ee
for all $\alpha\in\mathbb C$.
It follows immediately that (with $\epsilon = \epsilon_c = d_c(M,P)$)
\be
  |\langle 1|\bar M_1|2\rangle| 
    \leq \sum_{a}|\langle 1|F_{a1}|2\rangle|
    \leq 2\sqrt\epsilon+(N-2)\epsilon.
  \label{eq:me-upper-bound}
\ee
We now wish to find a lower bound for the same matrix element.

It is clear that
\be
  |\inprm{1}{2}|
    \geq |\inpr{1}{\bar 1}\inpr{\bar 1}{2}\inprm{\bar 1}{\bar 1}|-|A|-|B|-|C|-|D|,
  \label{eq:me-lower-bound}
\ee
with
\begin{subequations}
\begin{align}
  A &= \sum_{a \geq 2} \inpr{1}{\bar 1}\inpr{\bar a}{2}\inprm{\bar 1}{\bar a}, &
  B &= \sum_{a \geq 2} \inpr{1}{\bar a}\inpr{\bar 1}{2}\inprm{\bar a}{\bar 1},\\
  C &= \sum_{a \geq 2} \inpr{1}{\bar a}\inpr{\bar a}{2}\inprm{\bar a}{\bar a}, &
  D &= \sum_{\substack{a, b\geq 2 \\  a \neq  b}}
    \inpr{1}{\bar a}\inpr{\bar b}{2}\inprm{\bar a}{\bar b}.
\end{align}
\end{subequations}
To estimate $A$, let us consider the vector
\be
  |\psi\rangle = \delta e^{i\theta_{ 1}}|\bar 1\rangle
    +\sum_{a \geq 2}e^{i\theta_{a}}|\bar a\rangle,
\ee
with $\delta,\theta_a\in\mathbb R$, and the $\theta$s chosen such that
\be
  \inpr{\psi}{1} = \delta|\inpr{\bar 1}{1}|+\sum_{a \geq 2} |\inpr{\bar a}{1}|.
  \label{eq:A-ineq-1}
\ee
By Cauchy-Schwarz
\be
  \inpr{\psi}{1} \leq \|\psi\| = \delta^2 + N - 1.
  \label{eq:A-ineq-2}
\ee
Combining \eqref{eq:A-ineq-1} and \eqref{eq:A-ineq-2}, we find
\be
  \sum_{a \geq 2} |\inpr{\bar a}{1}| 
    \leq \sqrt{N-1}\sqrt{1-|\inpr{\bar 1}{1}|^2}.
\ee
With the abbreviations
\begin{align}
  t_a &= |\inpr{\bar 1}{a}|\text, & \tilde t_a &= \sqrt{1-t_a^2},
\end{align}
we thus have
\begin{subequations} \label{eq:ABCD-ineqs}
\begin{align}
  |A| &\leq \sqrt{N-1}t_1\tilde t_1 \sqrt{\bar \epsilon}, &
  |B| &\leq \sqrt{N-1}t_2\tilde t_2 \sqrt{\bar \epsilon},
\end{align}
and by similar considerations
\begin{align}
  |C| &\leq \tilde t_1\tilde t_2 \bar \epsilon, &
  |D| &\leq \sqrt{(N-1)(N-2)}\tilde t_1\tilde t_2 \bar \epsilon.
\end{align}
\end{subequations}
For mutual unbiased bases, it is simple to see that we can improve the estimate of $|D|$ to
\be
  |D| \leq \frac{(N-1)(N-2)}{N}\bar\epsilon.
\ee
From \eqref{eq:me-upper-bound}, \eqref{eq:me-lower-bound} and \eqref{eq:ABCD-ineqs}
it then follows that
\be
  I(\epsilon,\bar\epsilon;|\inpr{\bar 1}{1}|,|\inpr{\bar 1}{2}|) \geq 0,
  \label{eq:I-bound}
\ee
with
\begin{multline}
  I(\epsilon,\epsilon';t,t') = (N-2)\epsilon+2\sqrt\epsilon
    +\left(tt'+\tilde t \tilde t'[\sqrt{(N-1)(N-2)}+1]\right)\epsilon'\\
    +\sqrt{N-1}(t\tilde t+t'\tilde t')\sqrt{\epsilon'}
    -tt'.
\end{multline}
We thus have a non-trivial bound on $\epsilon$, $\bar\epsilon$ as long as
$|\inpr{\bar 1}{1}||\inpr{\bar 1}{2}| > 0$.

It is clear that we can combine the bounds \eqref{eq:I-bound} for different
combinations of basis vectors. Formally, let us define the subsets
\be
  H = \left\{(\epsilon,\bar\epsilon)\in[0,\infty)^2\middle|
    I(\epsilon,\bar\epsilon;|\langle \bar a|b\rangle|,|\langle \bar a|c\rangle|)
      \geq 0
    \text{ for all }
    a\text{ and }b\neq c\right\},
\ee
and
\be
  H' = \left\{(\epsilon,\bar\epsilon)\in[0,\infty)^2\middle|
    I(\bar\epsilon,\epsilon;|\langle a|\bar b\rangle|,|\langle a|\bar c\rangle|)
      \geq 0
    \text{ for all }
    a\text{ and }b \neq  c\right\}.
\ee
then we have
\be
  d_c(M,P) + d_c(\bar M,\bar P)
    \geq \inf\left\{\epsilon+\bar\epsilon \middle|(\epsilon,\bar\epsilon)\in H\cap H'\right\}.
\ee

\end{document}